# Chaotic Dynamics of Stellar Spin in Binaries and the Production of Misaligned Hot Jupiters

Natalia I. Storch[1], Kassandra R. Anderson[1], and Dong Lai[1]

[1]*Center for Space Research, Department of Astronomy, Cornell University, Ithaca, NY 14853*

**Many exoplanetary systems containing hot Jupiters are observed to have highly misaligned orbital axes relative to the stellar spin axes. Kozai-Lidov oscillations of orbital eccentricity/inclination induced by a binary companion, in conjunction with tidal dissipation, is a major channel for the production of hot Jupiters. We demonstrate that gravitational interaction between the planet and its oblate host star can lead to chaotic evolution of the stellar spin axis during Kozai cycles. As parameters such as the planet mass and stellar rotation period vary, periodic islands can appear in an ocean of chaos, in a manner reminiscent of other dynamical systems. In the presence of tidal dissipation, the complex spin evolution can leave an imprint on the final spin-orbit misalignment angles.**

About 1% of solar-type stars host giant planets with periods of ∼ 3 days (*1*). These "hot Jupiters" could not have formed in situ, given the large stellar tidal gravity and radiation fields close to their host stars. Instead, they are thought to have formed beyond a few astronomical units (AU) and migrated inward. However, the physical mechanisms of the migration remain unclear. In the last few years, high stellar obliquities have been observed in many hot Jupiter systems, i.e., the spin axis of the host star and the planetary orbital angular momentum axis are misaligned (*2–7*). Planet migration in protoplanetary disks (*8, 9*) is usually expected to produce aligned orbital and spin axes [however, see (*10–14*)], so the observed misalignment suggests that other formation channels may be required, such as strong planet-planet scatterings (*15, 16*), secular interactions/chaos between multiple planets (*17, 18*), and the Kozai-Lidov effect induced by a distant companion (*19–22*). Other observations suggest that multiple formation channels of hot Jupiters may be required (*23–25*).

In the "Kozai+tide" scenario, a giant planet initially orbits its host star at a few AU and experiences secular gravitational perturbations from a distant companion (a star or planet). When the companion's orbit is sufficiently inclined relative to the planetary orbit, the planet's eccentricity undergoes excursions to large values, while the orbital axis precesses with varying inclination. At periastron, tidal dissipation in the planet reduces the orbital energy, leading to inward migration and circularization of the planet's orbit.

As the planet approaches the star in a Kozai cycle, the planet-star interaction torque due to the rotation-induced stellar quadrupole makes the stellar spin and the planetary angular momentum axes precess around each other. Although the equations for such precession in the context of triple systems are known (*21, 26*), previous works on the "Kozai+tide" migration either neglected such spin-orbit coupling or included it without systematically examining the spin dynamics or exploring its consequences for various relevant parameter regimes (*19–22, 27*). However, the stellar spin has the potential to undergo rich evolution during the Kozai migration, which may leave its traces in the spin-orbit misalignments in hot Jupiter systems. Indeed, there are several examples of chaotic spin-orbit resonances in the Solar system. For instance, Saturn's satellite Hyperion experiences chaotic spin evolution due to resonances between spin and orbital precession periods (*28*). The rotation axis of Mars undergoes chaotic variation as well, as a result of resonances between the spin precession and a combination of orbital precession frequencies (*29, 30*).

We demonstrate here that gravitational interaction between the stellar spin and the planetary orbit can indeed induce a variety of dynamical behavior for the stellar spin evolution during Kozai cycles, including strongly chaotic behavior (with Lyapunov times as short as a few Myr) and perfectly regular behavior in which the stellar spin stays aligned with the orbital axis at all times. We show that in the presence of tidal dissipation the memory of chaotic spin evolution can be preserved, leaving an imprint on the final spin-orbit misalignment angles.

**Kozai Cycles and Spin-Orbit Coupling.** We consider a planet of mass $M_p$ initially in a nearly circular orbit around a star of mass $M_\star$ at a semi-major axis $a$, with a distant binary companion of mass $M_b$, semi-major axis $a_b$ and eccentricity $e_b$, which we set to 0. In that case, if the planet's initial orbital inclination relative to the binary axis, denoted by $\theta_{\rm lb}^0$, falls within the range $\{40°, 140°\}$, the distant companion induces cyclic variations in planetary orbit inclination and eccentricity, with a maximum eccentricity of $e_{\max} \simeq \sqrt{1 - (5/3)\cos^2\theta_{\rm lb}^0}$ (*31, 32*). These Kozai cycles occur at a characteristic rate given by

$$\begin{aligned} t_{\rm k}^{-1} &= n\left(\frac{M_b}{M_\star}\right)\left(\frac{a}{a_b}\right)^3 \\ &= \left(\frac{2\pi}{10^6 {\rm yr}}\right)\left(\frac{M_b}{M_\star}\right)\left(\frac{M_\star}{M_\odot}\right)^{1/2}\left(\frac{a}{1{\rm AU}}\right)^{3/2}\left(\frac{a_b}{100{\rm AU}}\right)^{-3}, \end{aligned} \quad (1)$$

where $n = 2\pi/P$ is the mean motion of the planet ($P$ is the orbital period). Note, however, that the presence of short-range forces, such as General Relativity and tidal distortions, tend to reduce the maximum attainable eccentricity, so that the actual $e_{\max}$ may be smaller than the "pure" (i.e. without short-range forces) Kozai value given above (*19, 20, 33*). Along with the eccentricity and inclination variations, the planet orbital angular momentum vector precesses around the binary axis ($\hat{\mathbf{L}}_b$) at an approximate rate which, in the absence of tidal dissipation, is given by (Sec. S1)

$$\Omega_{\rm pl} \approx \frac{3}{4} t_{\rm k}^{-1} \cos\theta_{\rm lb}^0 \sqrt{1 - e_0^2} \left[1 - 2\left(\frac{1 - e_0^2}{1 - e^2}\right)\frac{\sin^2\theta_{\rm lb}^0}{\sin^2\theta_{\rm lb}}\right], \quad (2)$$

where $e_0$ is the initial eccentricity. Because of the rotation-induced stellar quadrupole, the planet induces precession in the stellar spin orientation, governed by the equation

$$\frac{d\hat{\mathbf{S}}}{dt} = \Omega_{\text{ps}} \hat{\mathbf{L}} \times \hat{\mathbf{S}}. \tag{3}$$

Here $\hat{\mathbf{S}}$ and $\hat{\mathbf{L}}$ are unit vectors along the stellar spin and planet orbital angular momentum axes, respectively, and the precession frequency $\Omega_{\text{ps}}$ is given by

$$\begin{aligned} \Omega_{\text{ps}} &= -\frac{3GM_p(I_3 - I_1)}{2a^3(1-e^2)^{3/2}} \frac{\cos\theta_{\text{sl}}}{S} \\ &= -2.38 \times 10^{-8} \left(\frac{2\pi}{\text{yr}}\right) \frac{1}{(1-e^2)^{3/2}} \left(\frac{2k_q}{k_\star}\right) \left(\frac{10^3 M_p}{M_\star}\right) \left(\frac{M_\star}{M_\odot}\right)^{1/2} \left(\frac{\hat{\Omega}_\star}{0.1}\right) \left(\frac{a}{1\text{AU}}\right)^{-3} \left(\frac{R_\star}{R_\odot}\right)^{3/2} \cos\theta_{\text{sl}}, \end{aligned} \tag{4}$$

where $I_3$ and $I_1$ are principal moments of inertia of the star, $S$ is its spin angular momentum, $\hat{\Omega}_\star \equiv \Omega_\star/\sqrt{GM_\star/R_\star^3}$ is its spin frequency in units of the breakup frequency, $R_\star$ is the stellar radius, $\theta_{\text{sl}}$ is the angle between the stellar spin and planet angular momentum axes, and we have used $(I_3 - I_1) \equiv k_q M_\star R_\star^2 \hat{\Omega}_\star^2$ and $S \equiv k_\star M_\star R_\star^2 \Omega_\star$. For a solar-type star, $k_q \approx 0.05$, and $k_\star \approx 0.1$ (*34*). The stellar quadrupole also affects the planet's orbit, by introducing additional periastron advance, at a rate of order $-\Omega_{\text{ps}} S/(L\cos\theta_{\text{sl}})$ (where $L \equiv M_p\sqrt{GM_\star a(1-e^2)}$ is the orbital angular momentum), and making $\hat{\mathbf{L}}$ precess around $\hat{\mathbf{S}}$ at the rate $(S/L)\Omega_{\text{ps}}$ (Sec. S1).

During the Kozai cycle, orbital eccentricity varies widely from 0 to $e_{\text{max}}$, and thus $\Omega_{\text{ps}}$ and $\Omega_{\text{pl}}$ change from $\Omega_{\text{ps},0}$ and $\Omega_{\text{pl},0}$ to $\Omega_{\text{ps,max}}$ and $\Omega_{\text{pl,max}}$, respectively. However, $\Omega_{\text{ps}}$ is more sensitive to eccentricity variation than $\Omega_{\text{pl}}$, and attains a larger range of values. We therefore expect three qualitatively different regimes for the spin evolution.

Regime I, $|\Omega_{\text{ps,max}}| \lesssim |\Omega_{\text{pl,max}}|$ ("nonadiabatic"): $|\Omega_{\text{ps}}|$ is always smaller than $|\Omega_{\text{pl}}|$. We expect $\hat{\mathbf{S}}$ to effectively precess around $\hat{\mathbf{L}}_b$, the binary angular momentum axis (about which $\hat{\mathbf{L}}$ is precessing), maintaining an approximately constant angle $\theta_{\text{sb}}$.

Regime II, $|\Omega_{\text{ps,max}}| \gtrsim |\Omega_{\text{pl,max}}|$ and $|\Omega_{\text{ps},0}| \lesssim |\Omega_{\text{pl},0}|$ ("transadiabatic"): A secular resonance occurs when the stellar precession rate approximately matches the orbital precession rate ($|\Omega_{\text{ps}}| \approx |\Omega_{\text{pl}}|$). As the eccentricity varies from 0 to $e_{\text{max}}$ during the Kozai cycle, the system transitions from nonadiabatic to adiabatic. We expect this resonance crossing to lead to complex and potentially chaotic spin evolution.

Regime III, $|\Omega_{\text{ps},0}| \gtrsim |\Omega_{\text{pl},0}|$ ("adiabatic"): $|\Omega_{\text{ps}}|$ is always larger than $|\Omega_{\text{pl}}|$. We expect the spin axis to follow $\hat{\mathbf{L}}$ adiabatically, maintaining an approximately constant spin-orbit misalignment angle $\theta_{\text{sl}}$.

For a given planet semi-major axis $a$ and binary semi-major axis $a_b$, the division between different regimes depends on the product of planet mass and stellar spin (Fig. S1). In particular, systems with low $M_p$ and $\Omega_\star$ lie in Regime I, while those with large $M_p$ and $\Omega_\star$ lie in Regimes II and III.

**Numerical Exploration.** We first study the evolution of stellar spin in "pure" Kozai cycles: we integrate Eq. (3) together with the evolution equations for the planet's orbital elements, driven by the quadrupole potential from the binary companion (Sec. S1), but excluding all short-range forces. Although at the octupole level the companion may induce chaotic behavior in the planet orbit (*35–38*), the effect is negligible if $ae_b/[a_b(1-e_b^2)] \ll 0.01$ and is completely suppressed for $e_b = 0$. To isolate the dynamics of stellar spin evolution, we exclude the precession of $\hat{\mathbf{L}}$ around $\hat{\mathbf{S}}$ and all other short-range forces; thus, while the planet orbit influences the stellar spin, the stellar spin does not affect the orbit in any way. We consider different combinations of planet mass and stellar rotation rate to illustrate the different regimes described above (we consider $M_\star = M_b = 1 M_\odot$ and $R_\star = 1 R_\odot$ in all the examples shown in this paper). We present four "canonical" cases that encapsulate the range of the observed spin dynamics, including a sample trajectory in the transadiabatic regime (Regime II) (Fig. 1).

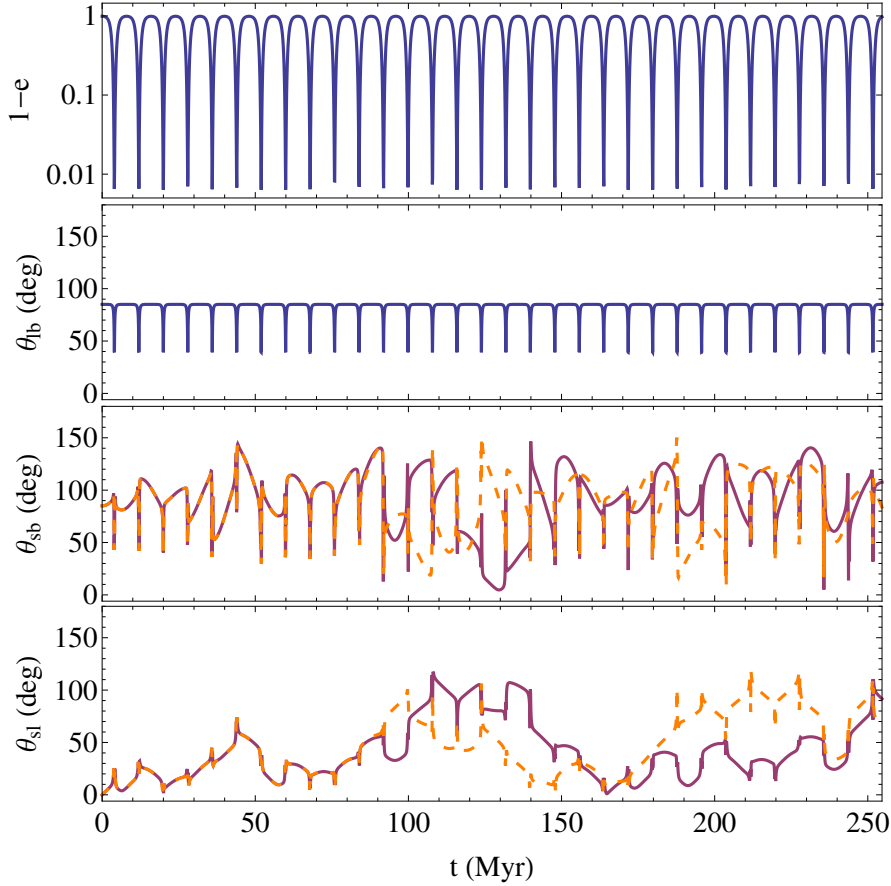

Figure 1: Sample evolution curves for the "pure" Kozai system, demonstrating how the stellar spin evolves through Kozai cycles. The parameters for this run are $a = 1\,\text{AU}$, $a_b = 200\,\text{AU}$, $e_b = 0$, $M_\star = M_b = 1 M_\odot$, $\tilde{\Omega}_\star = 0.05$, $M_p = 4.6 M_J$, and initial $e_0 = 0.01$, $\theta_{\text{lb}}^0 = 85°$. The spin's erratic evolution is suggestive of chaos; we therefore plot a "real" trajectory (red solid lines) and a "shadow" trajectory (orange dashed lines), used to evaluate the degree of chaotic behavior. The trajectories are initialized such that the "real" starts with $\hat{\mathbf{S}}$ parallel to $\hat{\mathbf{L}}$, and the "shadow" with $\hat{\mathbf{S}}$ misaligned by $10^{-6}\,\text{deg}$ with respect to $\hat{\mathbf{L}}$. This figure corresponds to the orange scatter plot of Fig. 2 and the orange curve of Fig. 3 (left). The spin evolution is highly chaotic.

We find excellent agreement with the qualitative arguments outlined above. In Regime I ("nonadiabatic", Fig. 2, top left) the spin evolution is regular and periodic. While we do not plot the spin-binary misalignment angle ($\theta_{\rm sb}$), it indeed stays constant. The "adiabatic" regime (Fig 2, bottom right) is difficult to access for trajectories that start with high initial misalignment of $\hat{\mathbf{S}}$ and $\hat{\mathbf{L}}$, due to the $\cos\theta_{\rm sl}$ factor in the spin precession frequency. Those trajectories that start with low initial $\theta_{\rm sl}$ (or with $\theta_{\rm sl}^0$ close to $180°$) maintain that angle, as expected. In Regime II ("transadiabatic"), two different types of behavior are observed. For most parameters that fall within this regime, the spin evolution is strongly chaotic, as indicated by the large degree of scatter that fills up the phase space (Fig. 2, top right and bottom left). However, periodic islands exist in the middle of this chaos, in which the stellar spin behavior is regular (Fig. 2, bottom left; Fig. S3).

Since the stellar spin and planet orbital axes in real physical systems typically start out aligned, we specialize to the trajectories with $\theta_{\rm sl}^0 = 0$ for the remainder of this paper. To assess the degree of chaos in each of the sample cases (Fig. 2), we evolve a "shadow" trajectory in addition to the real one (Fig. 1), with initial conditions very close to the original ones, and monitor how fast the two trajectories diverge, particularly in the spin direction. As expected, three out of four of our sample cases do not exhibit chaos, while the fourth, in the transadiabatic regime, is strongly chaotic, with a Lyapunov time of $\lambda^{-1} \sim 5.6$ Myr, corresponding to only $\sim 1$ Kozai cycle (Fig. 3, left).

Next we include the precession of $\hat{\mathbf{L}}$ about $\hat{\mathbf{S}}$ and other short range forces (periastron advances due to General Relativity, stellar quadrupole, planet's rotational bulge, and tidal distortion of the planet) (*19, 20*) in our calculations. We find that including these short-range forces for our four sample cases (Fig. 3, right) does not change our general conclusion that chaotic evolution occurs in the transadiabatic regime, although it can shift the locations (in the parameter space) of periodic islands.

Clearly, the stellar spin behavior in the transadiabatic regime is very complex: highly chaotic for certain parameters, more regular for others. To explore this diversity further, we construct a "bifurcation" diagram (Fig. 4), with which we could examine the degree of chaos over a large range of parameter values (particularly the planet mass). Visualized in this way, the topology of the chaos is more obvious: most of the mass bins are highly chaotic, but they are interspersed with individual, isolated quasiperiodic islands. To better understand this complex topology, we have developed a simpler analytical toy model that captures many of the features of this system

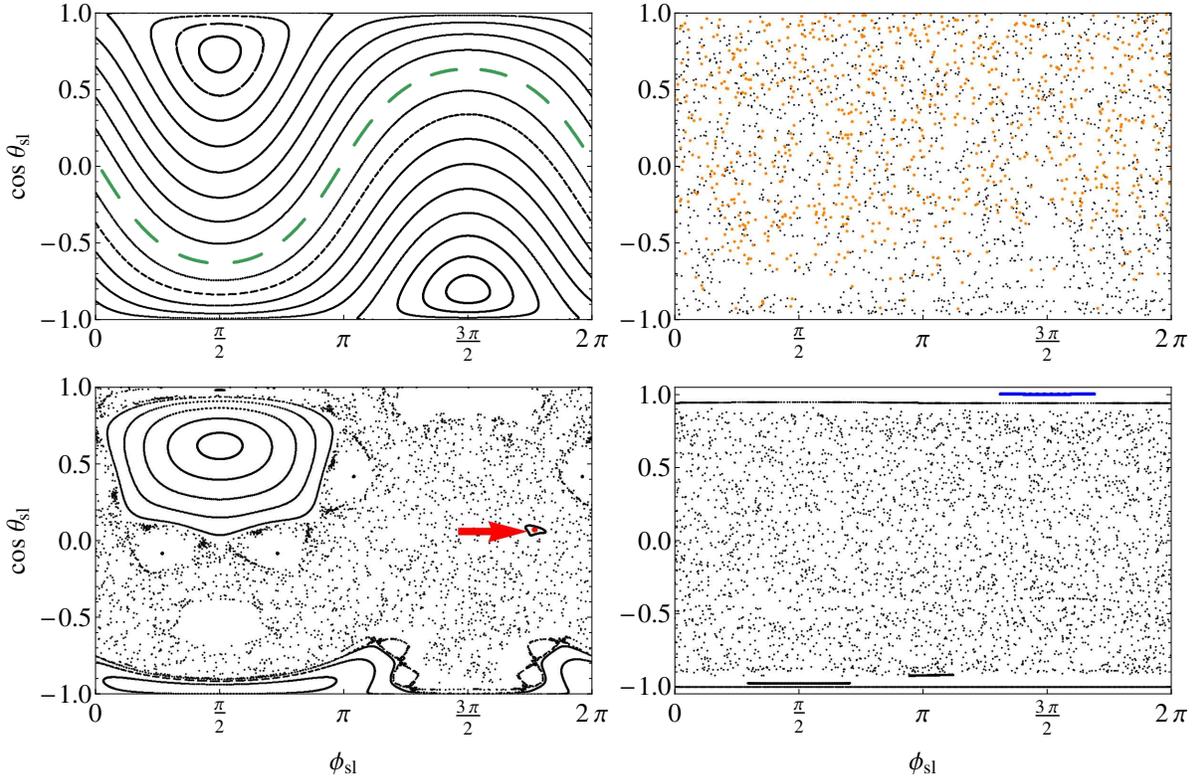

Figure 2: Surfaces of section of the angle ($\theta_{\rm sl}$) between $\hat{\bf S}$ and $\hat{\bf L}$ vs the precessional phase ($\phi_{\rm sl}$) of $\hat{\bf S}$ around $\hat{\bf L}$ for the "pure" Kozai system, demonstrating the presence or lack of chaos in the stellar spin evolution. In all of these sample cases, $a = 1\,{\rm AU}$, $a_b = 200\,{\rm AU}$, $e_b = 0$, $M_\star = M_b = 1 M_\odot$, and $e_0 = 0.01$, $\theta_{\rm lb}^0 = 85°$. Each panel is composed of multiple unique trajectories, corresponding to different initial $\theta_{\rm sl}^0$ (with the initial spin-binary angle $\theta_{\rm sb}^0$ ranging from 0 to $\pi$, and assuming $\hat{\bf S}$ is initially in the same plane as $\hat{\bf L}$ and $\hat{\bf L}_b$). In each panel the colored trajectory indicates the one with $\theta_{\rm sl}^0 = 0$. Each case is evolved for 12.7 Gyr, corresponding to $\sim 1500$ Kozai cycles. Each point in a trajectory is recorded at the argument of pericenter $\omega = \pi/2 (+ 2\pi n$, with $n$ an integer), corresponding to every other eccentricity maximum (Fig. S2). *Top left*: Regime I (nonadiabatic); $\hat{\Omega}_\star = 0.003$, $M_p = 1 M_J$. We show 18 unique trajectories, with $\theta_{\rm sb}^0$ ranging from 5° to 175°; the green line corresponds to $\theta_{\rm sl}^0 = 0$. The "equilibrium" states at $(\theta_{\rm sl}, \phi_{\rm sl}) \approx (40°, 90°)$ and $(40°, 270°)$ correspond to $\hat{\bf S}$ parallel and anti-parallel to $\hat{\bf L}_b$. *Top right*: Regime II (transadiabatic); $\hat{\Omega}_\star = 0.05$, $M_p = 4.6 M_J$; the orange dots show $\theta_{\rm sl}^0 = 0$, while the black dots are a composite of several different $\theta_{\rm sl}^0$. *Bottom left*: Regime II (transadiabatic); $\hat{\Omega}_\star = 0.03$, $M_p = 1.025 M_J$; 11 periodic or quasi-periodic trajectories and a composite chaotic region. The red dot at $(\cos\theta_{\rm sl}, \phi_{\rm sl}) \approx (0.06, 1.8\pi)$ (see arrow) corresponds to a periodic island with $\theta_{\rm sl}^0 = 0$. *Bottom right*: Regime III (onset of adiabaticity); $\hat{\Omega}_\star = 0.05$, $M_p = 20 M_J$; 5 quasi-periodic trajectories and a composite chaotic region. The blue line corresponds to $\theta_{\rm sl}^0 = 0$. Note that while both the orange and red cases are in Regime II, the orange one is highly chaotic, and the red resides in a periodic island.

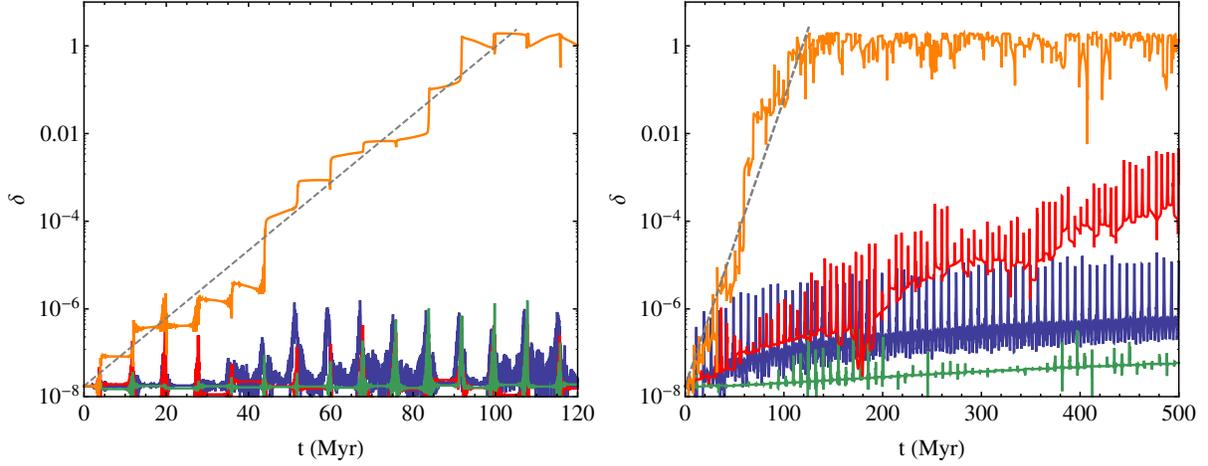

Figure 3: *Left panel*: Distance between two phase space trajectories, starting at slightly different initial spin orientations, for the "pure" Kozai system. The first (real) starts with $\hat{\mathbf{S}}$ parallel to $\hat{\mathbf{L}}$, the other (shadow) with $\hat{\mathbf{S}}$ misaligned by $10^{-6}$ deg with respect to $\hat{\mathbf{L}}$, for each of the sample $\theta_{sl}^0 = 0$ cases depicted in Fig. 2. The phase space distance is calculated as $\delta = |\hat{\mathbf{S}}_{\text{real}} - \hat{\mathbf{S}}_{\text{shadow}}|$ and therefore has a maximum value of 2. The lines are color-coded to correspond to each of the cases of Fig. 2. The grey dashed line demonstrates that for the chaotic orange curve, $\delta \propto e^{\lambda t}$, with $\lambda \sim 0.18\,\text{Myr}^{-1}$. *Right panel*: Same as left, but including orbit precession due to stellar quadrupole and periastron advances due to General Relativity, stellar quadrupole, planet oblateness, and static tides in the planet. The orange curve shows chaotic growth with $\lambda \sim 0.15\,\text{Myr}^{-1}$. The red curve, which is periodic on the left, is mildly chaotic here, with $\lambda \sim 0.02\,\text{Myr}^{-1}$.

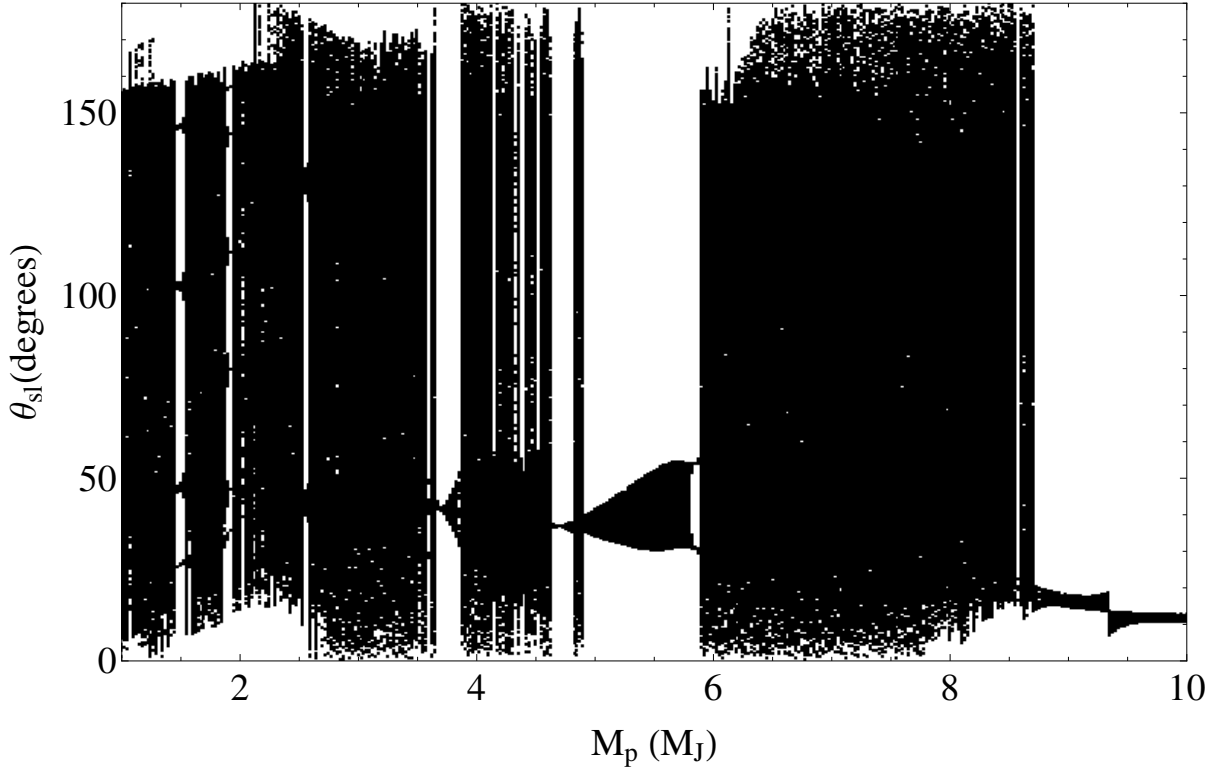

Figure 4: "Bifurcation" diagram of spin-orbit misalignment angle vs planet mass, including all short-range effects. The procedure described in Fig. 2 is carried out for each value of planet mass: the spin-orbit misalignment angle is recorded at every other eccentricity maximum for $\sim 1500$ Kozai cycles. The parameters for this plot are $a = 1\,\mathrm{AU}$, $a_b = 200\,\mathrm{AU}$, $e_0 = 0.01$, $\theta_{\mathrm{lb}}^0 = 85°$, $\hat{\Omega}_\star = 0.03$. High degree of scatter in a single mass bin indicates highly chaotic behavior. Note that multiple quasiperiodic islands appear in the middle of highly chaotic regions.

(Sec. S2.2).

Wide-spread chaos in dynamical systems is typically driven by overlapping resonances (*39*). Repeated secular spin-orbit resonance crossings ($|\Omega_{\mathrm{ps}}| \sim |\Omega_{\mathrm{pl}}|$) during Kozai cycles play an important role in producing the observed chaotic spin behavior. On the other hand, Kozai cycles themselves result from the near $1:1$ resonance ($\dot{\varpi} = \dot{\Omega}$) between the longitude of the periapse $\varpi$ and the longitude of the ascending node $\Omega$ of the planet's orbit. The back-reaction of the stellar spin on the orbit can naturally couple these two resonances. We suggest that all these effects are important in the development of the chaotic stellar spin evolution.

**Tidal dissipation and memory of chaotic evolution.** Having explored in some detail the variety of behaviors exhibited by stellar spin during Kozai cycles, we now assess the impact of this evolution on the production of hot Jupiters, particularly on their final stellar spin-orbit misalignment angles, by adding tidal dissipation to our equations. We employ the standard weak friction model of tidal dissipation in giant planets with constant tidal lag time (*40, 41*). In order to ensure that all our runs lead to circularized planets and a final $\theta_{\rm sl}$ within about $10^{10}$yrs, we enhance tidal dissipation by a factor of $14$ (Fig. 5, left) and $1400$ (Fig. 5, right) relative to the fiducial value for Jupiter (*42*) (Sec. S1). As long as the tidal evolution timescale of the orbit is much longer than the Lyapunov time for the chaotic spin evolution, we do not expect this enhancement to have major qualitative effect on the final observed spin-orbit misalignment angle.

Tidal dissipation leads to a gradual decrease in the proto-hot Jupiter's semi-major axis and eventual circularization close to the host star (Fig. S4). As the planet's orbit decays, Kozai cycles become suppressed by short-range forces. Also, as the semi-major axis decays, $|\Omega_{\rm ps}/\Omega_{\rm pl}|$ increases. Thus, even if we choose initial conditions that lie squarely in the nonadiabatic regime (Regime I), as $a$ decreases, all trajectories will eventually go through the $|\Omega_{\rm ps}| = |\Omega_{\rm pl}|$ secular resonance and end up fully adiabatic. At that point, the spin-orbit misalignment angle freezes out to some final, constant value $\theta_{\rm sl}^{\rm f}$.

In all of the numerical examples of non-dissipative evolution discussed above, we have held the value of the stellar spin rate $\Omega_\star$ constant. However, because the divisions between different spin evolution regimes depend on $\Omega_\star$, stellar spindown can potentially have a substantial effect on the degree of chaos in the system. Isolated solar-type stars spin down via magnetic braking associated with the stellar wind (*43*). For simplicity, we use the empirical Skumanich Law (*44*) to add stellar spindown to our evolution equations, starting with an initial spin period of $2.3$ days; the final spin period (at $t = 5$ Gyr) is $28$ days.

To assess the influence of chaotic stellar spin evolution on the final distribution of spin-orbit misalignment angles, we create a different kind of "bifurcation" diagram (Fig. 5). As in the non-dissipative case (Fig. 4), we consider a range of planet masses. For each $M_p$, we take a set of initial conditions that are identical in all but the initial orbit-binary misalignment angle $\theta_{\rm lb}$, which we randomly choose from a very small range: $\theta_{\rm lb}^0 \in \{86.99°, 87.01°\}$ (Fig. 5, left) and $\theta_{\rm lb}^0 \in \{84.95°, 85.05°\}$ (Fig. 5, right). We evolve these trajectories until the hot Jupiter circularizes and $\theta_{\rm sl}$ reaches its final value. We find that the scatter in $\theta_{\rm sl}^{\rm f}$ depends on the planet mass. The scatter generally increases with increasing $M_p$, but drops sharply in the adiabatic regime (for $M_p \gtrsim 4.4 M_J$ in the left panel of Fig. 5). There also exist quasiperiodic islands, where $\theta_{\rm sl}^{\rm f}$ has a rather small spread. Also, a range of misalignment angles around $90°$ appears to be excluded, with this range decreasing with increasing planet mass. Given the very small range of initial conditions, the evolution of any regular, non-chaotic system should result in only *one* final misalignment angle. Therefore, we suggest that this bimodality is the result of the system passing through the $|\Omega_{\rm ps}| \sim |\Omega_{\rm pl}|$ secular resonance, and the complex and possibly chaotic dynamics that occur during that time. We tentatively attribute the decrease of bimodality with increasing mass to an increase in chaotic behavior. The final semi-major axis $a_{\rm f}$ also exhibits "chaotic" spreads and periodic islands. Thus, in effect, the final distributions of $\theta_{\rm sl}^{\rm f}$ and $a_{\rm f}$ carry an imprint of the spin's past chaotic evolution.

As a final step, we run a "mini" population synthesis calculation, for a fixed value of $a_0$ and $a_b$ and a broader range of initial orbital inclinations (Fig. 6). A sharp contrast exists between the distribution of final spin-orbit misalignment angles at low $M_p$ and high $M_p$. At low $M_p$ a bimodal distribution of $\theta_{\rm sl}^{\rm f}$ is produced (this bimodality has been found in some previous popu-lation synthesis calculations (*20, 21*)). At high $M_p$ the evolution is mostly adiabatic, producing very little spin-orbit misalignment. This is a clear signature of the complex spin evolution in the observed stellar obliquity. Other factors, such as the stellar spindown rate and planetary tidal dissipation rate, can also affect the final misalignment distribution.

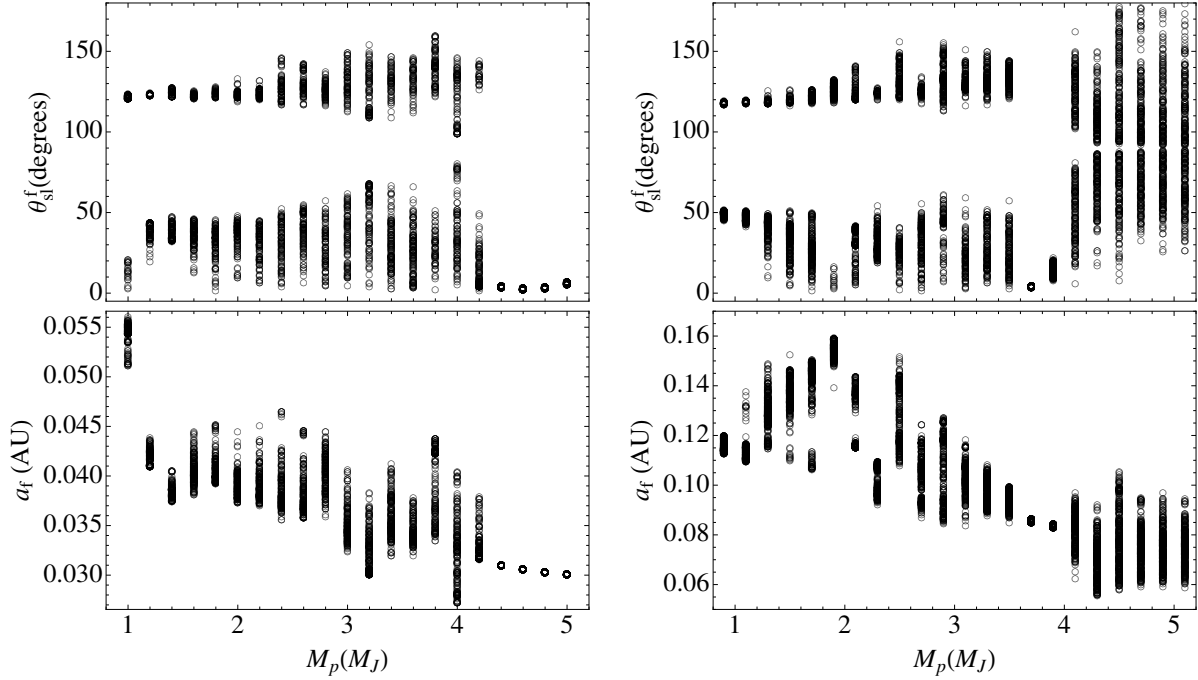

Figure 5: Two "bifurcation" diagrams of the final spin-orbit misalignment angle (top) and semi-major axis (bottom) vs planet mass for a small range of initial planet-binary inclinations, including the effects of tidal dissipation and stellar spindown. Here $a_b = 200\,\text{AU}$, $e_0 = 0.01$, $\hat{\Omega}_{\star,0} = 0.05$. Each data point represents the outcome of a single complete run starting with $a_0 = 1.5\,\text{AU}$ (left) and $a_0 = 1\,\text{AU}$ (right) and ending when the planet has sufficiently circularized (final eccentricity $e_\text{f} \leq 0.1$) and the final spin-orbit angle $\theta_\text{sl}^\text{f}$ is attained. For each run, we randomly select an initial inclination $\theta_\text{lb}^0$ from the range $86.99° - 87.01°$ (left) and $84.95° - 85.05°$ (right). Each mass bin contains $\sim 200$ points. The degree of scatter in $\theta_\text{sl}^\text{f}$ generally increases with increasing $M_p$, but drops sharply in the adiabatic regime (for $M_p \gtrsim 4.4 M_J$ in the left panel). Quasiperiodic islands are still present (e.g. at $\sim 3.8 M_J$ in the right panel).

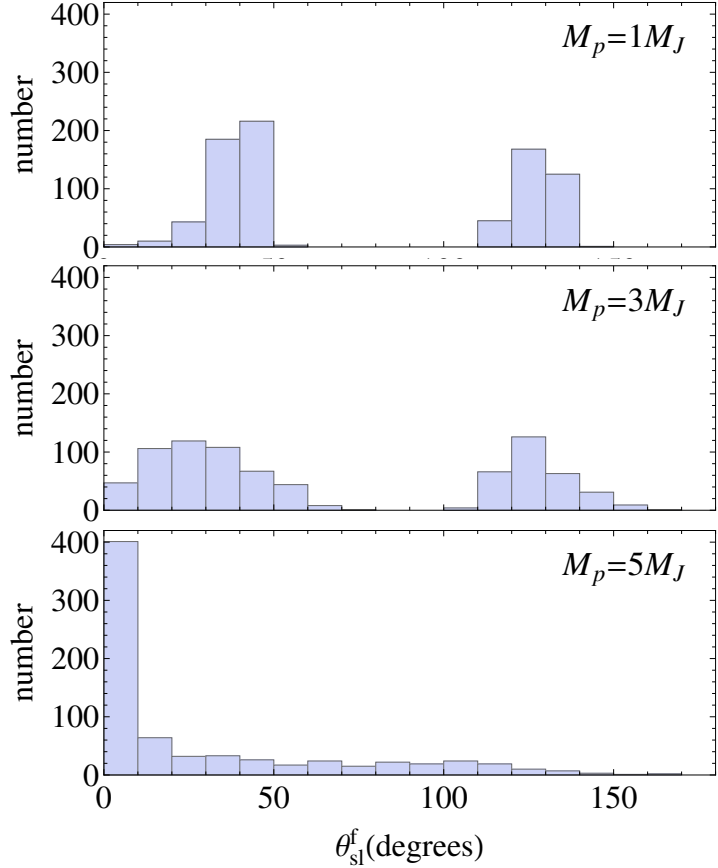

Figure 6: Distribution of the final spin-orbit misalignment angles as a function of planet mass, including the effects of tidal dissipation and stellar spin down, for initial planet-binary inclinations $\theta_{lb}^0$ in the range $85° - 89°$. Here $a_0 = 1.5\,\text{AU}$, $a_b = 200\,\text{AU}$, $e_0 = 0.01$, $\hat{\Omega}_{\star,0} = 0.05$. Each evolutionary trajectory is integrated until it has sufficiently circularized ($e_f \leq 0.1$), for a maximum of $5\,\text{Gyr}$. If by the end of $5\,\text{Gyr}$ the planet is not circularized, it is discarded. Note that the bimodality featured in Fig. 5 is still present here, despite the wider range of initial inclinations. At $M_p = 5M_J$ the evolution is mostly adiabatic, and therefore it is difficult to generate misalignment.

**Discussion.** The discovery of spin-orbit misalignment in close-in exoplanetary systems in the last few years was a major surprise in planetary astrophysics. Much of the recent theoretical work has focused on the non-trivial evolution of the planetary orbit (such as orbital flip) due to few-body gravitational interactions (*27, 36, 37*). However, as we have shown here, the spin axis of the host star can undergo rather complex and chaotic evolution, depending on the planetary mass and the stellar rotation rate. In many cases, the variation of the stellar spin axis relative to the binary axis is much larger than the variation of the orbital axis. Therefore, to predict the final spin-orbit misalignments of hot Jupiter systems in any high-eccentricity migration scenario, it is important to properly account for the complex behavior of stellar spin evolution.

In the above, we have focused on the Lidov-Kozai mechanism for the formation of hot Jupiters, but similar consideration can be applied to the formation of short-period stellar binaries (*20*). Indeed, spin-orbit misalignment angles have been measured for a number of close-in stellar binaries (*45–47*). Because of the much larger stellar spin precession rate in stellar binaries compared to the star-planet systems, the stellar spin evolution is expected to be largely in the adiabatic regime (depending on various parameters; Fig. S1), in which case the observed spin-orbit misalignment angles in close binaries would reflect their initial values at formation.

It is a curious fact that the stellar spin axis in a wide binary ($\sim 100$ AU apart) can exhibit such a rich, complex evolution. This is made possible by a tiny planet ($\sim 10^{-3}$ of the stellar mass) that serves as a link between the two stars: the planet is "forced" by the distant companion into a close-in orbit, and it "forces" the spin axis of its host star into wild precession and wandering.

The "binary+planet+spin" system studied in this paper exhibits many intriguing dynamical properties. While we have provided a qualitative understanding for the emergence of chaos in this system in terms of secular resonance crossing, much remains to be understood theoretically. Most remarkable is the appearance of periodic islands as the system parameters (planet mass and stellar spin) vary – a feature reminiscent of some well-known chaotic systems (*48, 49*).

50. We thank Konstantin Batygin, Dan Fabrycky, Matt Holman and Diego Muñoz for useful discussions. This work has been supported in part by NSF grants AST-1008245, AST-1211061 and NASA grants NNX12AF85G, NNX14AG94G. K.R.A. is supported by the NSF Graduate Research Fellowship Program under Grant No. DGE-1144153.

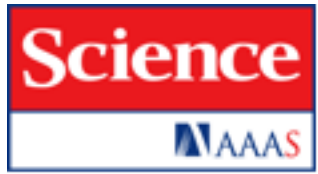

# Supplementary Materials for
Chaotic Dynamics of Stellar Spin in Binaries and the Production of Misaligned Hot Jupiters

Natalia I. Storch[1], Kassandra R. Anderson[1], and Dong Lai[1]

[1]*Center for Space Research, Department of Astronomy, Cornell University, Ithaca, NY 14853*

correspondence to: dong@astro.cornell.edu

**This PDF file includes:**
Materials and Methods
Supplementary Text
Figs. S1 to S7

# S1 Materials and Methods

For the "pure" Kozai problem discussed in the earlier part of the main text, we integrate the standard quadrupole Kozai-Lidov equations for the planet's orbital elements (assuming $M_p \ll M_\star, M_b$). These are given by

$$\begin{aligned}
\frac{de}{dt} &= t_{\rm k}^{-1} \frac{15}{8} e\sqrt{1-e^2} \sin 2\omega \sin^2 \theta_{\rm lb}, \\
\frac{d\Omega}{dt} &= t_{\rm k}^{-1} \frac{3}{4} \frac{\cos\theta_{\rm lb}\,(5e^2 \cos^2\omega - 4e^2 - 1)}{\sqrt{1-e^2}}, \\
\frac{d\theta_{\rm lb}}{dt} &= -t_{\rm k}^{-1} \frac{15}{16} \frac{e^2 \sin 2\omega \sin 2\theta_{\rm lb}}{\sqrt{1-e^2}}, \\
\frac{d\omega}{dt} &= t_{\rm k}^{-1} \frac{3\left[2(1-e^2) + 5\sin^2\omega(e^2 - \sin^2\theta_{\rm lb})\right]}{4\sqrt{1-e^2}},
\end{aligned} \qquad (S1)$$

where $e$ is the planet's orbital eccentricity, $\theta_{\rm lb}$ is the angle between the planet orbital angular momentum axis and the binary axis $\hat{\mathbf{L}}_b$, $\Omega$ is the longitude of the ascending node, $\omega$ is the argument of periastron, and $t_{\rm k}^{-1}$ is the characteristic Kozai rate, given by Eq. (1) of the main text. We choose the binary orbital plane to be the invariant plane. In all the cases we consider, we take as our initial condition $\Omega_0 = 0$ and $\omega_0 = 0$ (thus, $\omega$ always circulates rather than librates; see Fig. S2). Note, however, that this is not a particularly special choice, since for the initial inclinations $\theta_{\rm lb}$ we consider ($85° - 89°$) the maximum eccentricity is the same for the circulating and librating cases, and the rates of precession of the node ($\Omega_{\rm pl}$, Eq. 2) are only slightly different.

We evolve the precession of the stellar spin according to the equation

$$\frac{d\hat{\mathbf{S}}}{dt} = \Omega_{\rm ps} \hat{\mathbf{L}} \times \hat{\mathbf{S}}, \qquad (S2)$$

where $\Omega_{\rm ps}$ is given by Eq. (4), and $\hat{\mathbf{L}} = (\sin\theta_{\rm lb}\sin\Omega, -\sin\theta_{\rm lb}\cos\Omega, \cos\theta_{\rm lb})$ in the inertial frame where the $z$-axis is parallel to the binary axis $\hat{\mathbf{L}}_b$.

In the latter part of the main text, we add short-range forces to our system. We use the expressions given in *(19)* for periastron advances due to General Relativity, planet spin-induced quadrupole, and static tide in the planet. We also add nodal and apsidal precession of the planetary orbit due to the spin-induced stellar quadrupole. This introduces the following terms to the orbital evolution equations:

$$\begin{aligned}
\frac{d\omega}{dt} &= \omega_\star \left(1 - \frac{3}{2}\sin^2\theta_{\rm sl} - \frac{\cos\theta_{\rm lb}}{\sin\theta_{\rm lb}} \cos\theta_{\rm sl} \frac{\partial \cos\theta_{\rm sl}}{\partial \theta_{\rm lb}}\right), \\
\frac{d\Omega}{dt} &= \omega_\star \frac{\cos\theta_{\rm sl}}{\sin\theta_{\rm lb}} \frac{\partial \cos\theta_{\rm sl}}{\partial \theta_{\rm lb}}, \\
\frac{d\theta_{\rm lb}}{dt} &= -\omega_\star \frac{\cos\theta_{\rm sl}}{\sin\theta_{\rm lb}} \frac{\partial \cos\theta_{\rm sl}}{\partial \Omega},
\end{aligned} \qquad (S3)$$

where

$$\begin{aligned}
\cos\theta_{\rm sl} &= \hat{\mathbf{L}}\cdot\hat{\mathbf{S}} = S_x \sin\theta_{\rm lb}\sin\Omega - S_y \sin\theta_{\rm lb}\cos\Omega + S_z \cos\theta_{\rm lb}, \\
\frac{\partial\cos\theta_{\rm sl}}{\partial\theta_{\rm lb}} &= S_x \cos\theta_{\rm lb}\sin\Omega - S_y \cos\theta_{\rm lb}\cos\Omega - S_z \sin\theta_{\rm lb}, \\
\frac{\partial\cos\theta_{\rm sl}}{\partial\Omega} &= S_x \sin\theta_{\rm lb}\cos\Omega + S_y \sin\theta_{\rm lb}\sin\Omega,
\end{aligned} \quad (S4)$$

and $\omega_\star = -\Omega_{\rm ps} S/(L\cos\theta_{\rm sl})$.

Finally, we add tidal dissipation in the planet to our equations. We use the standard weak friction tidal dissipation model *(40,41)*:

$$\frac{1}{a}\frac{da}{dt} = \frac{1}{t_a}\frac{1}{(1-e^2)^{15/2}}\left[(1-e^2)^{3/2}f_2(e)\frac{\Omega_{\rm s,p}}{n} - f_1(e)\right], \quad (S5)$$

$$\frac{1}{e}\frac{de}{dt} = \frac{11}{4}\frac{1}{t_a}\frac{1}{(1-e^2)^{13/2}}\left[(1-e^2)^{3/2}f_4(e)\frac{\Omega_{\rm s,p}}{n} - \frac{18}{11}f_3(e)\right], \quad (S6)$$

where $a$ is the semi-major axis, $\Omega_{\rm s,p}$ is the spin rate of the planet, the functions $f_1 - f_4$ are defined as

$$\begin{aligned}
f_1(e) &= 1 + \frac{31}{2}e^2 + \frac{255}{8}e^4 + \frac{185}{16}e^6 + \frac{25}{64}e^8, \\
f_2(e) &= 1 + \frac{15}{2}e^2 + \frac{45}{8}e^4 + \frac{5}{16}e^6, \\
f_3(e) &= 1 + \frac{15}{4}e^2 + \frac{15}{8}e^4 + \frac{5}{64}e^6, \\
f_4(e) &= 1 + \frac{3}{2}e^2 + \frac{1}{8}e^4,
\end{aligned} \quad (S7)$$

$$(S8)$$

and $t_a$ is a characteristic timescale, given by

$$\frac{1}{t_a} = 6k_2 \Delta t_{\rm L}\left(\frac{M_\star}{M_p}\right)\left(\frac{R_p}{a}\right)^5 n^2, \quad (S9)$$

where $n$ is the mean motion of the planet, $k_2$ is the tidal Love number and $\Delta t_{\rm L}$ is the tidal lag time. For Jupiter, $k_2 = 0.37$ and we take $\Delta t_{\rm L} = 0.1$ s (corresponding to $k_2/Q \approx 10^{-5}$ at a tidal forcing period of 6.5 hours). We therefore use $\Delta t_{\rm L} = 0.1\chi$ s, where $\chi$ is a tidal enhancement factor, which we take to be 14 for Fig. 5 (left) and 1400 for Fig. 5 (right), in order to ensure that the planets in our test cases circularize within the lifetime of their host stars. For all the sample cases considered in this work, we assume the planet spin to be pseudosynchronous with the orbit, i.e. $\Omega_{\rm s,p}/n = f_2(e)/[(1-e^2)^{3/2}f_5(e)]$, with $f_5(e) = 1 + 3e^2 + (3/8)e^4$. Relaxing this assumption does not qualitatively change our results. (For pseudosynchronous spin, the periastron advance due to planet's rotation bulge is always smaller than that due to tidal distortion.)

Equivalent evolution equations for the spin-triple system can be found in *(21,26)*.

# S2 Supplementary Text

## S2.1 Figures

In this section we provide several supplementary figures that facilitate deeper understanding of the rich dynamics exhibited by the stellar spin during Kozai cycles and migration.

As stated in the main text, the division between different regimes of stellar spin behavior depends on the planet semi-major axis, binary semi-major axis, and the product of planet mass and stellar spin frequency. In Fig. S1, we illustrate these divisions in the $a_b - a$ space for several different values of $\hat{M}_p \equiv (\hat{\Omega}_\star/0.05)(M_p/M_J)$. We note that for real systems, short-range effects due to General Relativity (GR) and tidal/rotation distortion of the planet may affect the Kozai cycles. For the parameter space explored in this paper, the GR effect dominates. When the Kozai precession frequency $\dot{\omega}_k \sim t_k^{-1}(1-e^2)^{3/2}$ becomes comparable to the GR-induced precession frequency $\dot{\omega}_{GR}$, the Kozai cycle is arrested. In this case, the maximum eccentricity achieved during a Kozai cycle is reduced, and any planet undergoing Kozai cycles in will fail to become a hot Jupiter if $r_p = a(1 - e_{max})$ is larger than $\sim 0.1$ AU. Thus, the effect of GR can restrict the available parameter space in which adiabatic evolution (regime III) happens *and* a hot Jupiter is created. However, the presence of short-range forces and tidal dissipation also alters the topology of the chaos in the parameter space, making it difficult to draw a direct connection between the regime divisions in the "pure" Kozai system and the results of our dissipative simulations. In fact, the results of Fig. 5 (left) demonstrate that, indeed, it is possible for hot Jupiters to experience adiabatic evolution.

In order to explore the three regimes of stellar spin evolution, we create surfaces of section (Fig. 2) by sampling the spin trajectory every time the orbital trajectory comes back to the same region of phase space. In Fig. S2 we show the orbital trajectory in phase space, with and without short-range forces, and mark the point at which we choose to sample the spin evolution.

In the main text, we demonstrate that in the "transadiabatic" regime (regime II), stellar spin has the potential to undergo both chaotic motion and regular quasiperiodic motion, depending on the parameters of the system. In Fig. 1 we present an example of a chaotic trajectory. Here, in Fig. S3 we present an example of a periodic transadiabatic trajectory: even at late times, the "real" and "shadow" trajectories match perfectly.

Finally, in Fig. S4 we present a sample time evolution for the Kozai problem with added short-range forces, tidal dissipation and stellar spindown, showing how the final semi-major axis $a_f$ and spin-orbit misalignment angle $\theta_{sl}^f$ are attained. Each point in Fig. 5 represents the result of such evolution.

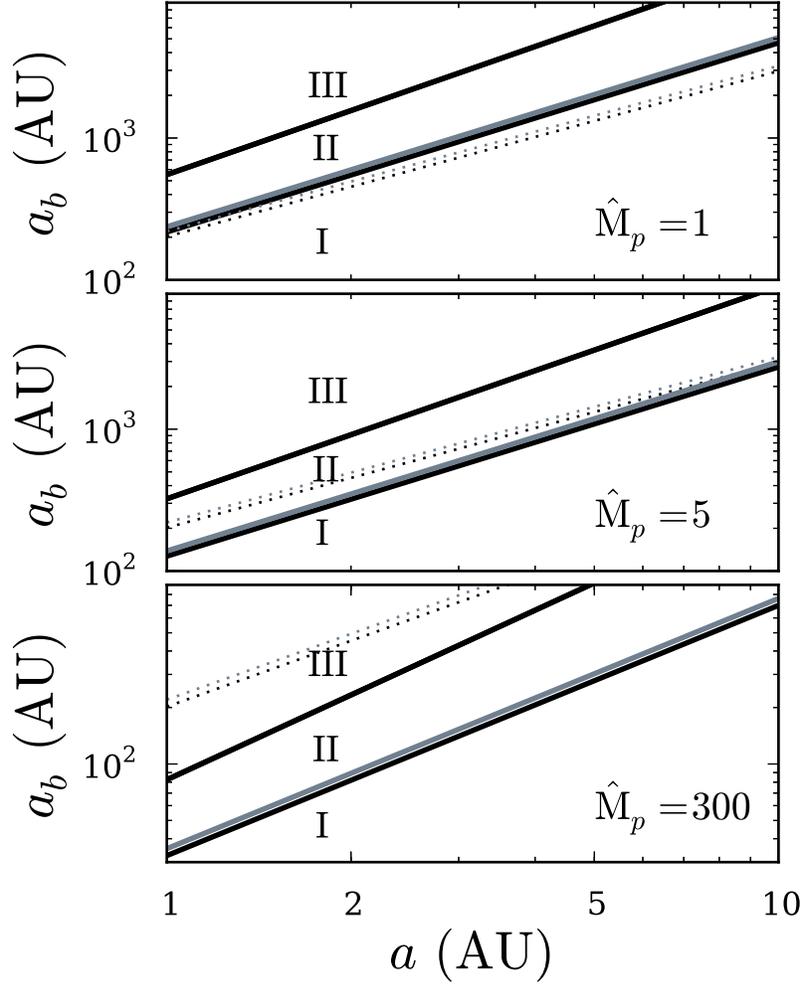

Figure S1: Breakdown of parameter space into the three regimes of spin evolution, as discussed in the text. *Black*: for a periastron distance of $r_p = a(1 - e_{\max}) = 0.03$ AU; *gray*: for $r_p = 0.05$ AU. Here $\hat{M}_p = (\hat{\Omega}_\star/0.05)(M_p/M_J)$. The regimes are determined by the relative values of the stellar spin precession frequency $\Omega_{\rm ps}$ and the nodal precession frequency $\Omega_{\rm pl}$ of the planet's orbit. Note that $\Omega_{\rm ps}$ depends on $\cos\theta_{\rm sl}$, and for concreteness we use $\cos\theta_{\rm sl} = 1$. $\Omega_{\rm pl}$ is a complicated function of eccentricity and $\theta_{\rm lb}$ (Eq. 2), which we approximate as $\Omega_{\rm pl} \approx -t_{\rm k}^{-1}/(1 - e^2)$ in making this figure. The lines separating Regimes I and II are given by $|\Omega_{\rm ps,max}| \approx 0.5|\Omega_{\rm pl,max}|$, where $\Omega_{\rm ps,max}$ and $\Omega_{\rm pl,max}$ are equal to $\Omega_{\rm ps}$ and $\Omega_{\rm pl}$ evaluated at $(1-e_{\max}) = r_p/a$. The line separating Regimes II and III is given by $|\Omega_{\rm ps,0}| \approx 2|\Omega_{\rm pl,0}|$, where $\Omega_{\rm ps,0}$, $\Omega_{\rm pl,0}$ are equal to $\Omega_{\rm ps}$ and $\Omega_{\rm pl}$ evaluated at $e = 0$. The dotted lines mark the boundary at which the effect of GR becomes significant, approximated by $\dot{\omega}_{\rm GR} \approx t_{\rm k}^{-1}(1 - e_{\max}^2)^{-1/2}$. Above the dotted lines, GR will suppress the Kozai cycles, so that the system cannot reach the specified $r_p$. In Regimes I and III the spin precession frequency never overlaps with the nodal precession frequency, and the spin evolution is expected to be regular and periodic. In Regime II, the two frequencies are equal for some value of $e$ during the Kozai cycle, and therefore secular spin-orbit resonance develops, potentially leading to chaos. Note that the parameters shown in the lowest panel ( $\hat{M}_p = 300$) correspond to a low-mass star rather than a planet.

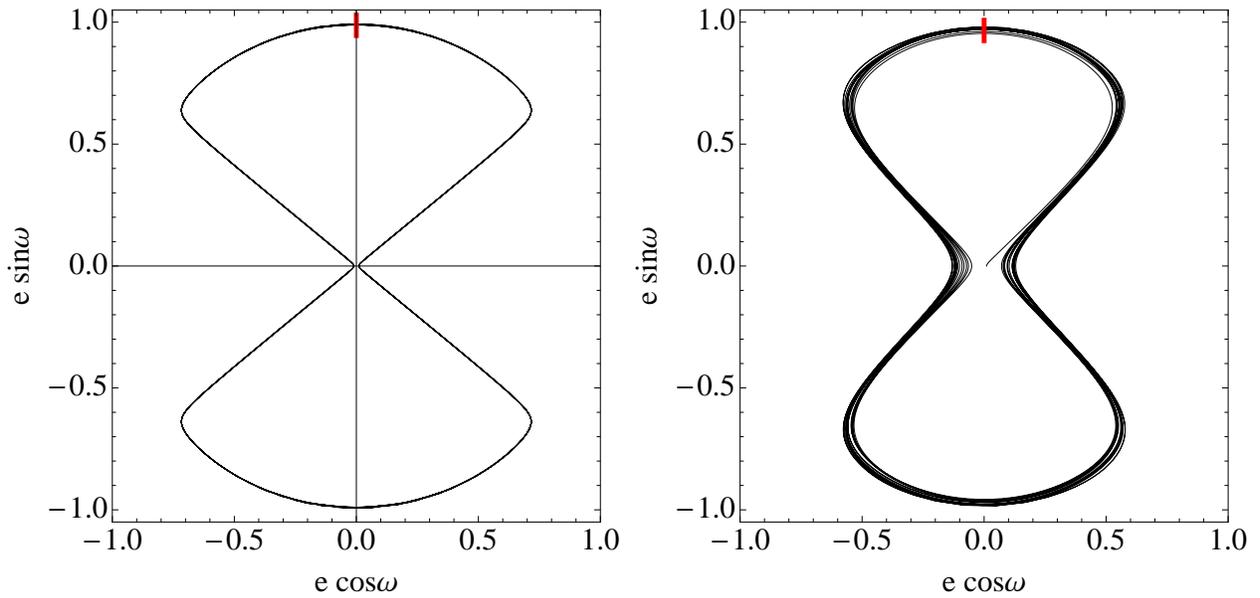

Figure S2: Orbital trajectory in $e - \omega$ phase space, for the "pure" Kozai problem (left), and with the addition of short-range forces (right). $\omega$ circulates with a period that is twice the period of the eccentricity oscillations. In red, we mark the point in the trajectory where we choose to sample the spin evolution in generating Figs. 2 and 4: i.e., every time the trajectory passes that point, we record the stellar spin orientation.

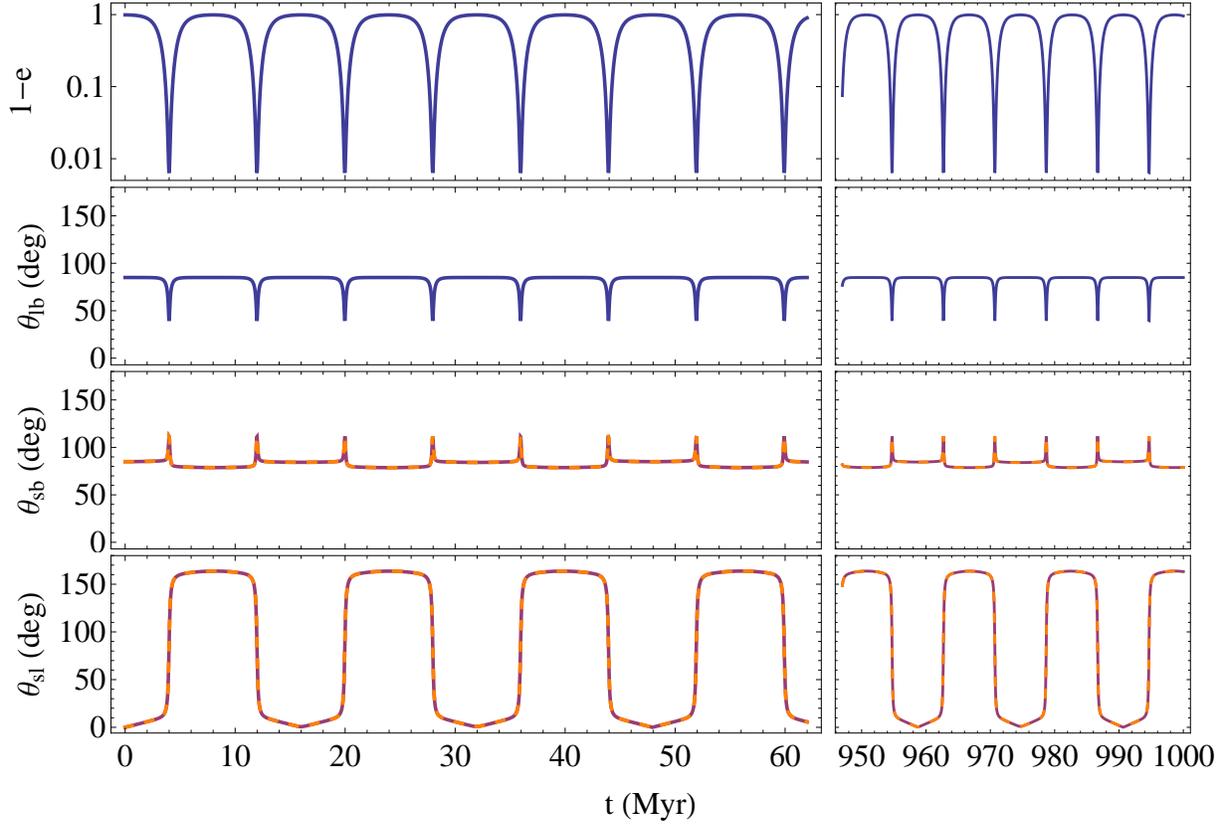

Figure S3: Sample evolution curves for a trajectory in a periodic island of regime II, demonstrating how the stellar spin evolves through many Kozai cycles. We plot a "real" trajectory (red solid lines) and a "shadow" trajectory (orange dashed lines), used to evaluate the degree of chaotic behavior. The trajectories are initialized such that the "real" starts with $\hat{\mathbf{S}}$ parallel to $\hat{\mathbf{L}}$, and the "shadow" with $\hat{\mathbf{S}}$ misaligned by $10^{-6}$deg with respect to $\hat{\mathbf{L}}$. The parameters are $a = 1\text{AU}$, $a_b = 200\text{AU}$, $e_0 = 0.01$, $\theta_{\text{lb}}^0 = 85°$, $\hat{\Omega}_\star = 0.03$, $M_p = 1.025 M_J$. This figure corresponds to the red points of Fig. 2 (bottom left) and the red curve of Fig. 3 (left). It is perfectly periodic: even at late times, the "real" and "shadow" trajectories match perfectly.

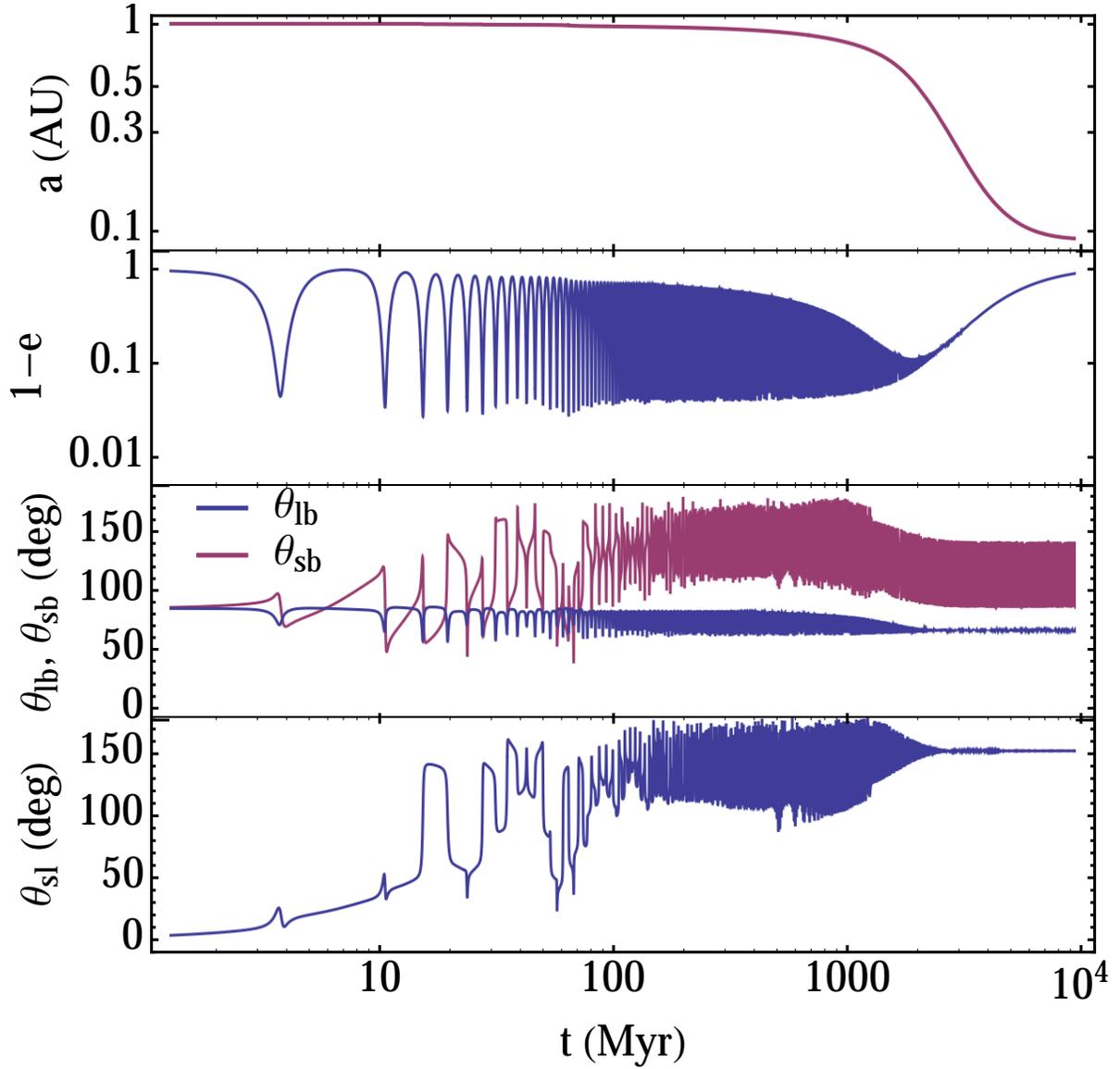

Figure S4: Sample orbital and spin evolution, including tidal dissipation and stellar spindown. The parameters for this run are $a_0 = 1\text{AU}$, $a_b = 200\text{AU}$, $e_0 = 0.01$, $\theta_{\text{lb}}^0 = 85°$, $\hat{\Omega}_{\star,0} = 0.05$, $M_p = 5M_J$, $\chi = 700$.

## S2.2 Toy Model

We consider a toy model in order to gain a better understanding of the dynamical behavior of the "real" Kozai system with stellar spin evolution (i.e. the system on which we focused in the main text). In this model, the stellar spin axis $\hat{\mathbf{S}}$ satisfies Eq. (S2), and the orbital axis $\hat{\mathbf{L}}$ evolves according to

$$\frac{d\hat{\mathbf{L}}}{dt} = \Omega_{\text{pl}} \hat{\mathbf{L}}_b \times \hat{\mathbf{L}}, \tag{S10}$$

where we have neglected the back-reaction torque of the stellar spin on the planetary orbit (this back-reaction can be included but it does not introduce qualitatively new features when $L \gg S$), and the nutation of the orbital angular momentum vector $\hat{\mathbf{L}}$. The external binary axis $\hat{\mathbf{L}}_b$ is fixed in time, and the angle between $\hat{\mathbf{L}}$ and $\hat{\mathbf{L}}_b$ is constant. The spin precession rate $\Omega_{\text{ps}}$ is a function of eccentricity (and time) [see Eq. (4)]. In the case of pure Kozai oscillations (i.e. without extra precession effects), the eccentricity is a periodic function of time, varying between $0$ and $e_{\max}$. We imitate this oscillatory behavior by adopting the following explicit form for $\Omega_{\text{ps}}$:

$$\Omega_{\text{ps}}(t) = \Omega_{\text{ps},0} f(t) \cos\theta_{\text{sl}}, \quad \text{with} \quad f(t) \equiv \frac{1+\varepsilon}{1+\varepsilon \cos\Omega_0 t}, \tag{S11}$$

where $\Omega_0$ represents the Kozai oscillation frequency. The precession frequency of $\hat{\mathbf{L}}$ around $\hat{\mathbf{L}}_b$ has the approximate eccentricity dependence $\Omega_{\text{pl}} \propto [2(1-e^2)^{-1} - 1]$ in the real system, and therefore in our toy model takes the form

$$\Omega_{\text{pl}} = \Omega_{\text{pl},0}(2f^{2/3} - 1), \qquad \text{where} \qquad \Omega_{\text{pl},0} = \frac{3}{4}\Omega_0 \cos\theta_{\text{lb}}. \tag{S12}$$

During a Kozai cycle, $\Omega_{\text{ps}}$ varies from $\Omega_{\text{ps},0}\cos\theta_{\text{sl}}$ to $\Omega_{\text{ps,max}} = \Omega_{\text{ps},0}(1+\varepsilon)\cos\theta_{\text{sl}}/(1-\varepsilon)$. We adopt $\varepsilon = 0.99$ in our examples below. Thus, the parameter $\omega_{\text{ps},0} \equiv \Omega_{\text{ps},0}/\Omega_{\text{pl},0}$ determines whether the system is nonadiabatic ($\omega_{\text{ps},0} \lesssim 0.1$), transadiabatic ($0.1 \lesssim \omega_{\text{ps},0} \lesssim 1$), or fully adiabatic ($\omega_{\text{ps},0} \gtrsim 1$).

For a given $\Omega_{\text{ps},0}$, we numerically integrate Eqs. (S2) and (S10) for 1000 "Kozai cycles," record the values of $\theta_{\text{sl}}$ and $\theta_{\text{sb}}$ at eccentricity maxima (i.e., $\Omega_0 t = \pi, 3\pi, 5\pi, \cdots$), and then plot these values in the $\theta_{\text{sl}} - \omega_{\text{ps},0}$ and $\theta_{\text{sb}} - \omega_{\text{ps},0}$ planes. We repeat the process for different values of $\omega_{\text{ps},0}$. The results are shown in Fig. S5 for initial $\theta_{\text{lb}} = 60°$ (and initial $\theta_{\text{sl}} = 0°$). The range of $\omega_{\text{ps},0}$ has been chosen to illustrate the nonadiabatic, transadiabatic and fully adiabatic regimes.

As in the real system, our toy model exhibits periodic/quasiperiodic solutions and chaotic zones, and the level of chaos is determined by the parameter $\omega_{\text{ps},0}$. If we use the spreads of $\theta_{\text{sl}}$ and $\theta_{\text{sb}}$ as a measure of chaos, we see that the system generally becomes more chaotic with increasing $\omega_{\text{ps},0}$, until $\omega_{\text{ps},0}$ reaches $\sim 5$, beyond which the system becomes fully-adiabatic ($\theta_{\text{sl}} \to 0$ and $\theta_{\text{sb}}$ approaches a constant). However, multiple periodic islands exist in the ocean of chaos. Figure S6 illustrates the time evolution of $\theta_{\text{sl}}$ and $\theta_{\text{sb}}$ in several of these periodic islands, along with an example of chaotic evolution. Figure S7 compares $\delta(t) = |\hat{\mathbf{S}}_{\text{real}}(t) - \hat{\mathbf{S}}_{\text{shadow}}(t)|$ (where the shadow trajectory has an initial condition nearly identical to the real one) for the different cases, clearly showing the difference between the periodic islands and chaotic evolution.

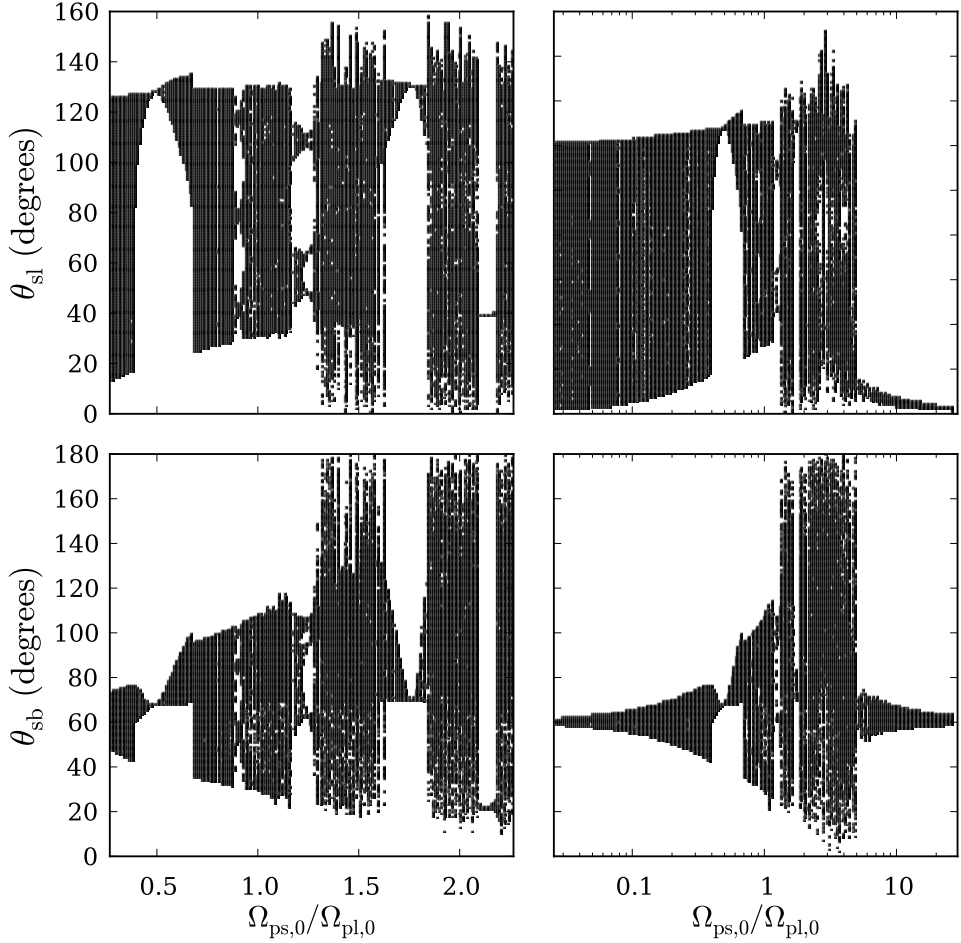

Figure S5: Angles $\theta_{\rm sl}$ and $\theta_{\rm sb}$ evaluated at maximum eccentricity (where $\Omega_0 t = \pi, 3\pi, 5\pi...$ for 1000 cycles) as functions of $\omega_{\rm ps,0} \equiv \Omega_{\rm ps,0}/\Omega_{\rm pl,0}$. The initial angle between $\hat{\mathbf{L}}$ and $\hat{\mathbf{L}}_b$ is $\theta_{\rm lb}^0 = 60°$, and $\hat{\mathbf{S}}$ and $\hat{\mathbf{L}}$ are initially aligned. The range of $\omega_{\rm ps,0}$ (on the logarithmic scale) in the right panels is chosen to illustrate the behavior of the three regimes (nonadiabatic, transadiabatic, and fully adiabatic). The narrow range of $\omega_{\rm ps,0}$ (on the linear scale) in the left panels exhibits the existence of periodic and quasiperiodic islands within the (chaotic) transadiabatic zones.

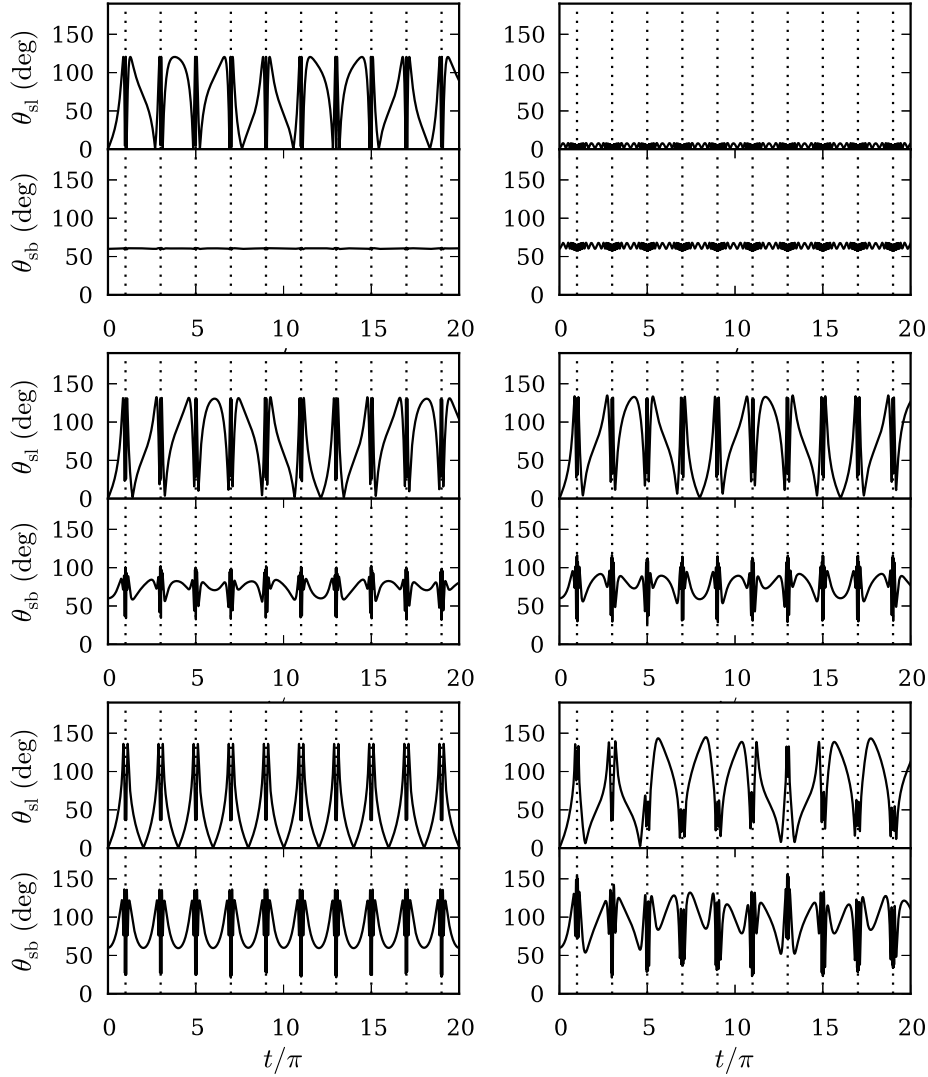

Figure S6: Angles $\theta_{\rm sl}$ and $\theta_{\rm sb}$ as functions of time, demonstrating the various behaviors of different orbits shown in Figure S5, including the three distinct regimes, and the difference between periodic and chaotic evolution in the transadiabatic regime. Time is in units of $\Omega_0 = 1$ (Eq. S11), and has been scaled by $\pi$. The dashed lines, included for reference, are located at odd-integers (when the system is at maximum eccentricity). *Upper left panel*: $\omega_{\rm ps,0} \equiv \Omega_{\rm ps,0}/\Omega_{\rm pl,0} = 0.023$, nonadiabatic, so that $\theta_{\rm sb} \approx$ constant. *Upper right panel*: $\omega_{\rm ps,0} = 13.3$, fully adiabatic, so that $\theta_{\rm sl} \approx \theta_{\rm sl}^0 \approx 0$. *Middle left panel*: $\omega_{\rm ps,0} = 0.89$, transadiabatic but periodic, with period= $12\pi$. *Middle right panel*: $\omega_{\rm ps,0} = 1.25$, transadiabatic but periodic, with period= $16\pi$. *Bottom left panel*: $\omega_{\rm ps,0} = 2.13$, transadiabatic but periodic, with period= $2\pi$. *Bottom right panel*: $\omega_{\rm ps,0} = 2.35$, transadiabatic, with no discernible periodic behavior, chosen to illustrate chaotic evolution. See also Fig. S7 for further comparison between periodic and chaotic evolution.

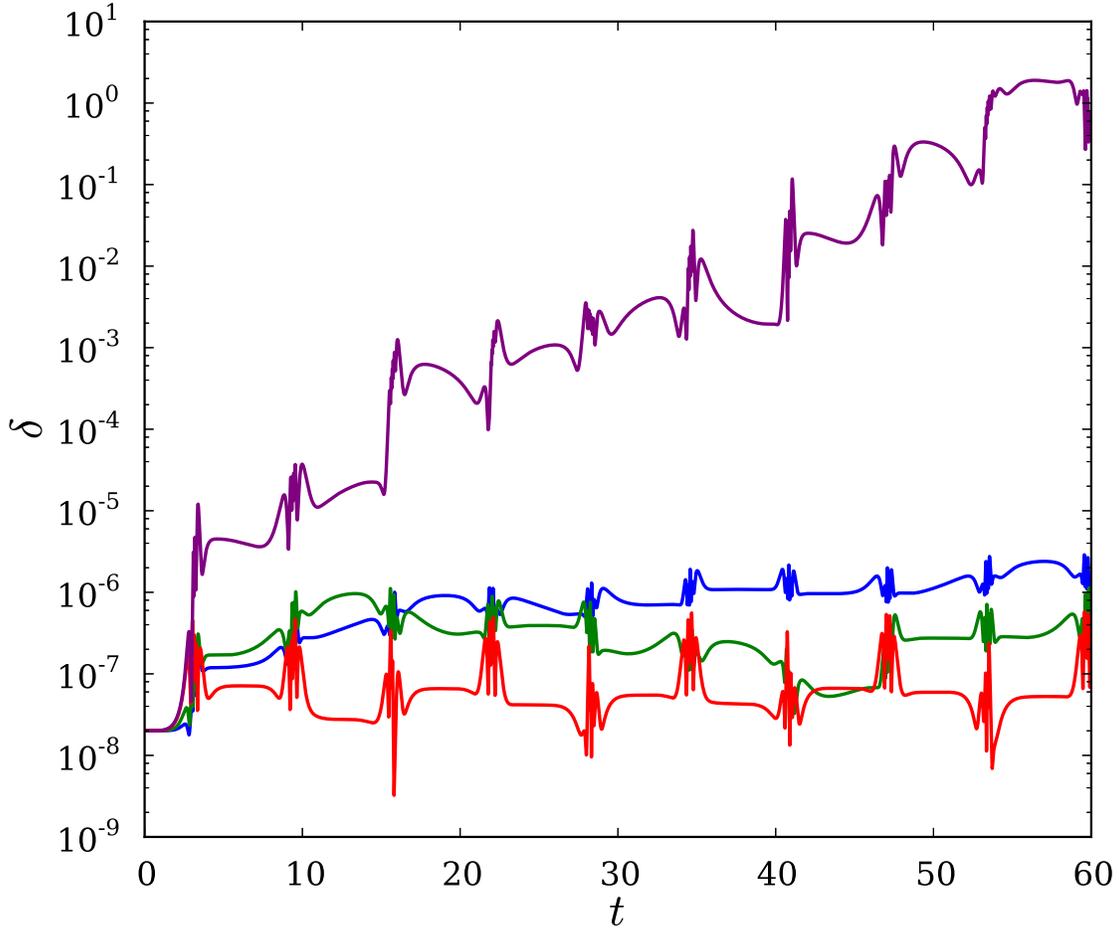

Figure S7: Difference ($\delta$) in the spin vector $\hat{\mathbf{S}}$ between "real" and "shadow" trajectories for the four transadiabatic systems shown in Fig. S6 (bottom 4 panels), starting with an initial $\delta_0 = 10^{-8}$. Time is in units of $\Omega_0 = 1$. Three examples of periodic evolution are shown, where $\omega_{\text{ps},0} \equiv \Omega_{\text{ps},0}/\Omega_{\text{pl},0} = 0.89$ (blue), $\omega_{\text{ps},0} = 1.25$ (green), $\omega_{\text{ps},0} = 2.13$ (red), as well as a chaotic example $\omega_{\text{ps},0} = 2.35$ (purple). Compare with Figure S6. For the periodic examples $\delta$ remains small, while in the chaotic example, $\delta$ increases exponentially, and eventually saturates to its maximum value of $\delta = 2$.